# A New Method to Predict the Effect of an Intervention in the Host Population to Reduce the Magnitude of an Outbreak of a Vector-Borne Infection

**Francisco Antônio Bezerra Coutinho[1], Marcos Amaku[2], Esper Georges Kallas[2,3] and Eduardo Massad[3,4]**

1. Instituto de Física, Universidade de São Paulo, São Paulo, Brasil
2. Faculdade de Medicina, Universidade de São Paulo, São Paulo, Brasil
3. Instituto Butantan, São Paulo, Brasil
4. Faculdade de Saúde Pública, Universidade de São Paulo, São Paulo, Brasil

## Abstract

In this paper, we propose a new model to estimate the impact of an intervention on human hosts of a vector-borne infection, such as dengue, which occurs in yearly outbreaks of different magnitudes. The model applies to these outbreaks and, in fact, is independent of their intensity, that is, it does not require the steady-state assumption. The model takes as input the officially reported age-dependent number of cases of a vector-borne infection. It is deterministic and does not account for stochasticity. Our objective is to estimate the impact of the intervention (the efficacy), and we rely on the observed fact that the age distribution of the proportion of cases of the infections transmitted by the same vector is independent of both the intensity of transmission and the geographic area studied, at least for Brazilian regions. This finding is highlighted in the main text and forms the basis of our calculations. A hypothetical intervention is simulated using a dengue vaccine, which allows the determination of the optimal strategy for a vaccination campaign.

## Introduction

The estimation of the efficacy of control measures against infectious diseases has been the subject of numerous studies. For example, for directly transmitted infections, Anderson and May (1983) studied measles and rubella, assuming the dynamic evolution of a time perturbation around a steady state. Hethcote (1988) proposed the analysis of a steady-state vaccination strategy against measles based on a fixed age-dependent force of infection. Massad et al. (1994) applied Hethcote's model to a rubella vaccine and incorporated the effects of several pulse vaccination campaigns.

When the steady-state assumption is relaxed, the analysis becomes considerably more complex. See, for example, Amaku et al. (2003), who studied vaccination against rubella and measles in several countries. This study, as well as Anderson and May (1983), analyzed the time evolution of infections following the introduction of a vaccine. However, Anderson and May (1983) assumed a constant, age-independent contact rate between individuals, whereas Amaku et al. (2003) considered several age-dependent contact functions, all consistent with observed seroprevalence data. Both studies concluded that the transition toward a new equilibrium is very slow and depends on the vaccination schedule. Amaku et al. (2003) further concluded that the optimal vaccination strategy against rubella in Brazil consists of a mass vaccination campaign followed by the incorporation of the vaccine into the routine immunization schedule.

For vector-borne infections, Massad et al. (2005) applied Bauch's (2003) game-theoretical model to optimize a vaccination strategy against yellow fever, balancing the risk of outbreaks against the risk of adverse events. These risks were calculated using a Ross–Macdonald dynamic model, with outbreaks simulated numerically from the introduction of a single infected vector into a constant mosquito population. In another study, Amaku, Coudeville, and Massad (2012) assumed steady-state transmission and evaluated several vaccination strategies against dengue.

As noted above, most studies addressing age-dependent infection assume steady-state conditions, i.e., no temporal variation. However, vector-borne infections typically exhibit annual outbreaks driven by seasonal variation in vector populations (Coutinho et al., 2006; Amaku et al., 2015, 2016; Massad et al., 2017; Coutinho et al., 2025).

When the steady-state assumption is relaxed, standard mathematical methods face substantial challenges. Moreover, predicting vector population dynamics and other factors that may trigger outbreaks is extremely difficult (Coutinho et al. 2006; Amaku et al. 2015, 2016). For example, vector populations depend strongly on climatic factors and exhibit marked seasonality (Coutinho et al. 2006; Pinto et al. 2011). In addition, the occurrence of outbreaks in human populations depends on the movement of infected individuals within cities (Amaku et al., 2016).

Indeed, in a deterministic framework without the steady-state assumption, the equations describing transmission dynamics between mosquito populations and human hosts take the form of partial differential equations with unknown time-dependent parameters, such as vector population size and human mobility patterns (Anderson and May, 1983; Amaku et al., 2015, 2016).

In this paper, we show that for vector-borne infections, it is possible to develop a formalism that compares the magnitude of an outbreak in the absence of intervention with that observed under an intervention (e.g., vaccination). The model assumes that the intervention does not alter mosquito biting behavior or human behavior toward mosquitoes (Boccia et al., 2014; Massad et al., 2016; Ferreira et al., 2017).

Using this framework, we propose a new method to estimate the impact of intervention strategies against vector-borne infections outside the steady-state assumption, and importantly, independent of temporal variations in transmission intensity. We illustrate the method using dengue data from the city of São José do Rio Preto (SJRP) in the State of São Paulo, South Eastern Brazil, covering the outbreak of 2019 and studied the efficacy of several vaccination strategies against dengue with a recently licensed vaccine against dengue (Butantan-DV in Peixoto de Miranda et al. 2025).

## Methods and Results

### The data

The dengue data were obtained from the National System of Compulsory Notification of Infections of the Brazilian Ministry of Health (SINAN) (SINAN-MOH 2025), which provides the dengue incidences stratified by ages for each year.

### The model

Assume an outbreak of a vector-borne infection beginning at the start of the transmission season, $t = 0$ (September in the Southern Hemisphere), and ending at the end of transmission season, $t = T$ (August of the next year in the Southern Hemisphere).

Denote the reported double density number of new cases (incidence) of individuals with ages between $a$ and $a + da$ occurring between the times $t$ and $t + dt$ as $Inc_H(a, t)$. Then, the total number of cases per outbreak ending at a specific time $T$, $\Lambda_T$, (that is, the end of the outbreak that occurs at a given season) is $Inc_H(a, T)$ summed over all ages, and in $t$ from 0 to $T$, and is given by:

$$\Lambda_T = \int_0^T \left[\int_0^\infty Inc_H(a, t) da\right] dt = \int_0^T \Lambda(t) dt \qquad (1)$$

Using equation (1), we can write:

$$Inc_H(a,t) = \Lambda_T \frac{Inc_H(a,t)}{\int_0^\infty \int_0^T Inc_H(a,t)dadt} \quad (2)$$

The fraction $\frac{Inc_H(a,t)dt}{\int_0^\infty \int_0^T Inc_H(a,t)dadt}$ in equation (2) represents the proportion of reported cases between $t$ and $t + dt$, with respect to the total number of cases, $\Lambda_T$, occurring in individuals aged between $a$ and $a + da$. We observed that this quantity is approximately a function of age $a$ only, independently of both location and the year (Coutinho et al., 2026) and is denoted:

$$\phi(a) = \frac{\int_0^T Inc_H(a,t)dt}{\int_0^\infty \int_0^T Inc_H(a,t)dadt} \quad (3)$$

It is remarkable that, for the sites and the years analyzed, this function is independent of both the magnitude of the outbreak and its geographical location, whether within a municipality or across distant regions of the country. In fact, in a previous paper (Coutinho et al., 2026), we show that the distribution of the proportion of reported cases of aedes-transmitted infections is approximately the same for Brazil as a whole, Rio de Janeiro city and São José do Rio Preto city for all outbreaks occurring between 2001 and 2024 (Figure 1).

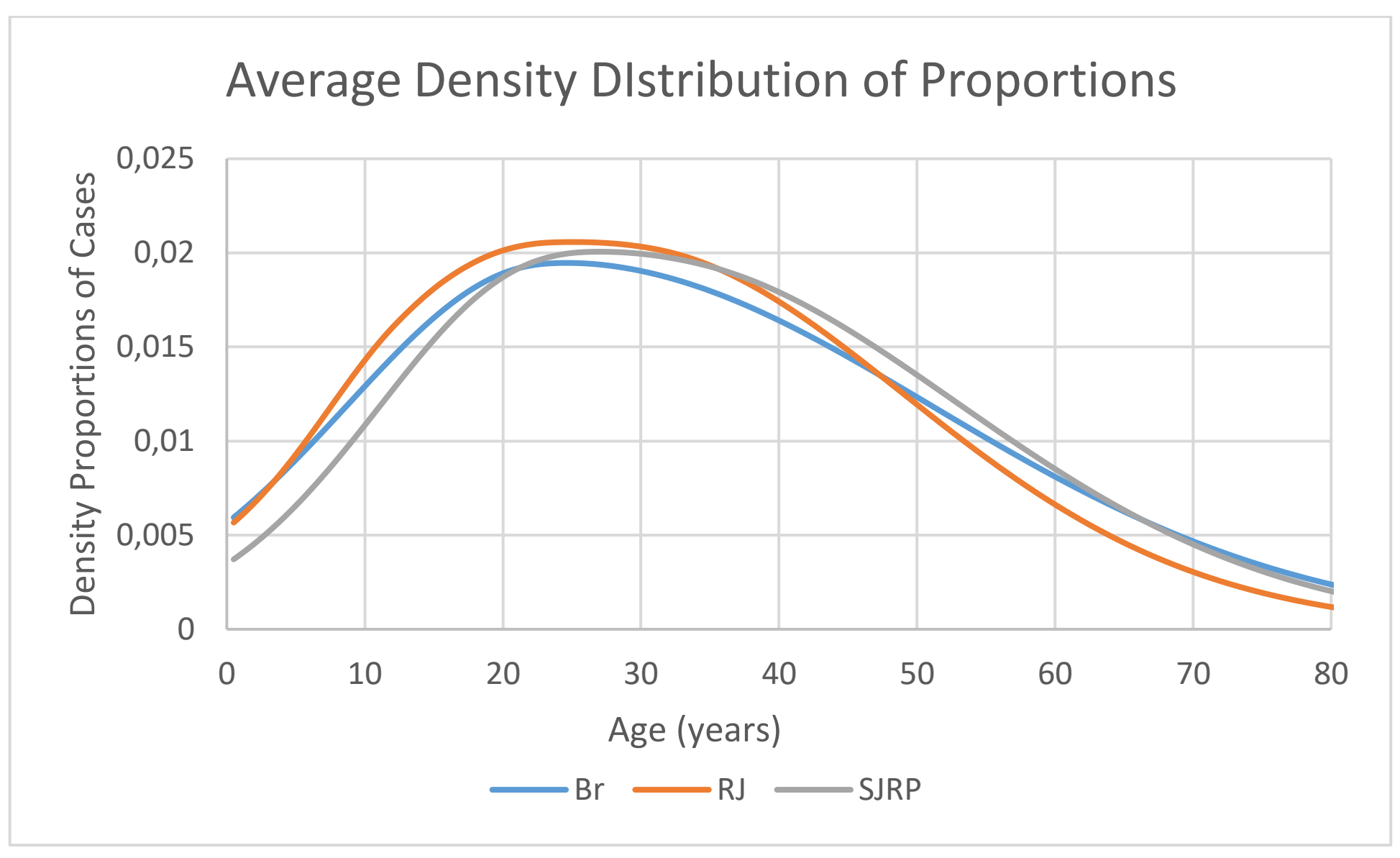

**Figure 1. Average age distribution of the proportion of dengue cases in Brazil as a whole, Rio de Janeiro city and São José do Rio Preto city (from Coutinho et al., 2026).**

We hypothesize (Coutinho et al., 2026) that this is due to the exposure of the host population to the vectors and this is valid only for urban centers, because the vector studied is an urban species (Guzman and Kouri, 2002; Cecilio et al., 2009; Powell et al, 2013). For infections transmitted by different vectors, like malaria or leishmaniases, the age-distribution of the proportion of cases is entirely different.

Now, for vector-borne infections, the yearly incidence, $\Lambda_T$, is a function of the product of the total number of infected mosquitoes during the outbreak (twelve months season) which ends at $T$. Now, let $\mathcal{S}_H(t)dt$ be the number of susceptible humans between the instants $t$ and $t + dt$. Therefore, the total number of susceptible humans involved in the outbreak up to time $t$ is

$$\mathcal{S}_{Ht} = \int_0^t \mathcal{S}(s)ds \tag{4}$$

or, in the total outbreak

$$\mathcal{S}_{HT} = \int_0^T \mathcal{S}_H(s)ds \qquad (5)$$

Hence, the total proportion of susceptible humans is $\mathscr{s}_{HT} = \frac{\mathcal{S}_{HT}}{\mathcal{N}_{HT}}$. $\mathcal{S}_{HT}$ and $\mathcal{N}_{HT}$ are the total number of susceptible individuals and the total human population, respectively, involved in the outbreak.

In addition, let $\mathcal{I}_{Mt}$ be the total number of infected mosquitoes up to time $t$, defined as

$$\mathcal{I}_{Mt} = \int_0^t \imath_M(s)ds \qquad (6)$$

where $\imath_M(t)dt$ in the number of infected mosquitoes which acquired the infection between $t$ and $t + dt$.

The total number of infected mosquitoes at the end of the outbreak is

$$\mathcal{I}_{MT} = \int_0^T \imath_M(t)dt \qquad (7)$$

Therefore, since when any of the variables $\mathcal{S}_{HT}$ and $\mathcal{I}_{MT}$ vanishes, the incidence vanishes, it is natural to think that the total incidence is a function of the product of $\mathcal{S}_{Ht}$ and $\mathcal{I}_{Mt}$. Denoting the incidence $Inc_H(t)$ as a function $k'f((\mathcal{S}_{Ht}\,\mathcal{I}_{Mt})$, where $k'$ is a constant with dimension $time^{-1}$, and expanding in Maclaurin series, and keeping only the first non-zero term, we have

$$Inc_H(t) = k'f(\mathcal{S}_{Ht}\,\mathcal{I}_{Mt}) = k'\frac{\partial f(0)}{\partial(\mathcal{S}_{H0}\,\mathcal{I}_{M0})}\mathcal{S}_{Ht}\,\mathcal{I}_{Mt} + \cdots \qquad (8)$$

Calling $k = k'\frac{\partial f(0)}{\partial(\mathcal{S}_{H0}\,\mathcal{I}_{M0})}$ we have that the number of human cases between $t$ and $t + dt$, $\Lambda(t)$ is, in first order

$$\mathrm{d}\Lambda(t) = k\mathcal{S}_{Ht}\,\mathcal{I}_{Mt}dt \tag{9}$$

Equation (8) corresponds (but it is not the same) to Macdonald's (1952) notation, in the context of malaria, the incidence of cases in humans. In Macdonald's notation, the human incidence $\lambda_H(t)$, reads:

$$\lambda_H(t) = abI_M(t)\frac{S_H(t)}{N_H(t)} \tag{10}$$

where $a$ is the mosquitoes' biting rate, $b$ is the probability of transmission from vectors to hosts and $I_M(t)$ and $\frac{S_H(t)}{N_H(t)}$ are point prevalence.

**Simulating an intervention**

Let us now assume a given control strategy, e.g. a vaccine, whose effect is 'removing' a fraction $p$ of reported cases in the age interval between $a_1$ and $a_2$ and that the intervention occurred during the outbreak ending at $T$. As mentioned before, since the proportion of cases, $\phi(a)$, is independent of the year and location (Coutinho et al., 2026), we define $\phi_c(a, a_1, a_2, p)$ as the proportion of human cases resulting from the intervention, which depends on the fraction $p$ and $a_1$ and $a_2$ only (see Figure 3).

In order to illustrate the above calculation, we simulated the result of a vaccination against dengue in São José do Rio Preto in 2019, covering 100% of the entire population in the age interval between 10 and 20-year-olds. Figure 2 shows the historical series of dengue outbreaks in SJRP from 2014 to 2024. We choose the 2019 outbreak because it is clear and very strong.

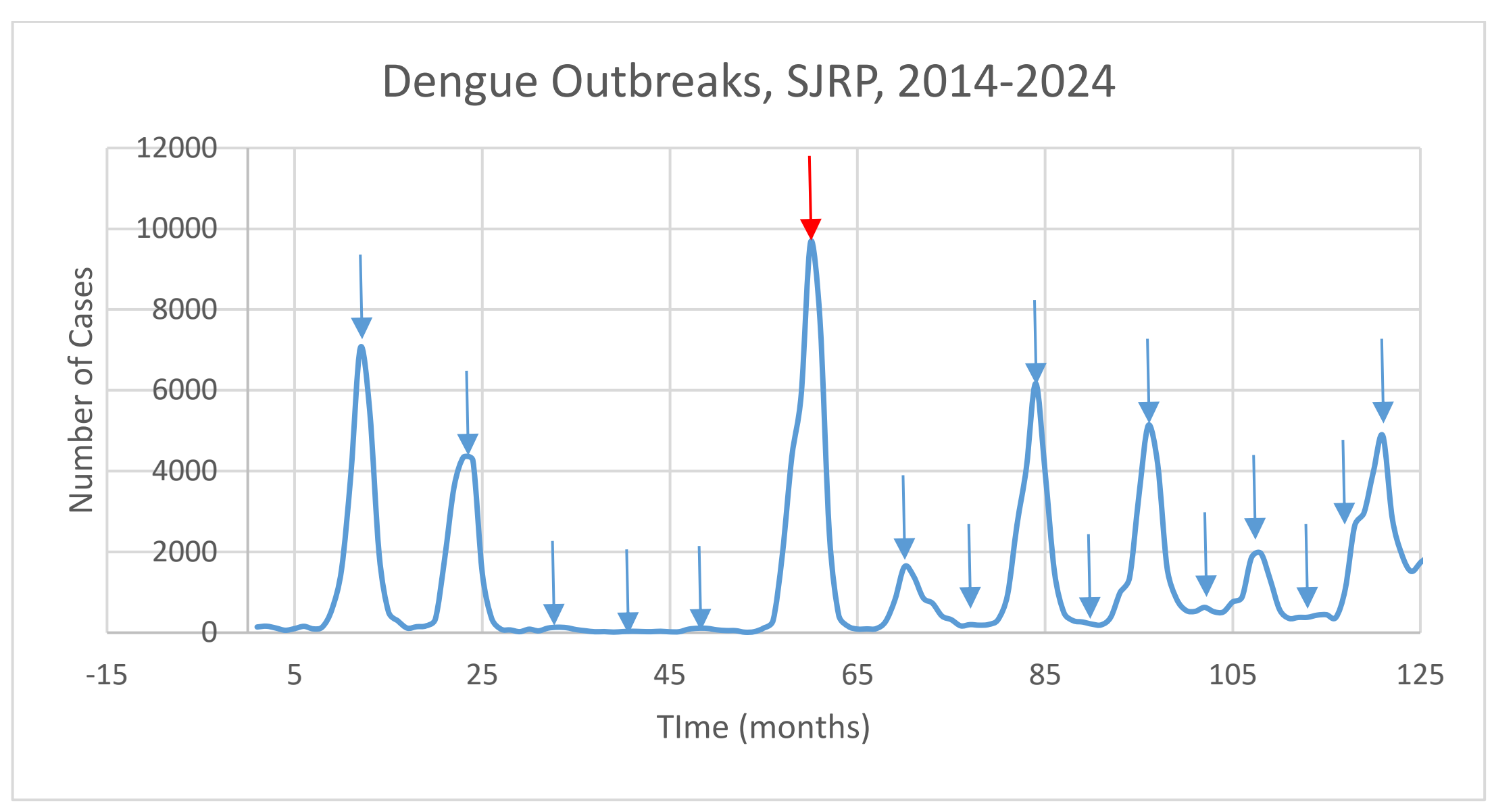

**Figure 2. Dengue outbreaks of dengue in São José do Rio Preto city for the period 2014-2024. Small arrows indicate the peak of each yearly outbreak. The red arrow points to the 2019 outbreak (Coutinho et al., 2026). The 2019 outbreak is preceded by three small outbreaks, barely seen in the figure.**

As seen in Figure 2, there are outbreaks for every yearly transmission season (arrows). It is very difficult to predict the size of a yearly transmission season outbreak, that is, if it will be a big or a small outbreak. However, our method works for any outbreak size since it compares the result of the intervention for any outbreak with what would be the same outbreak without intervention.

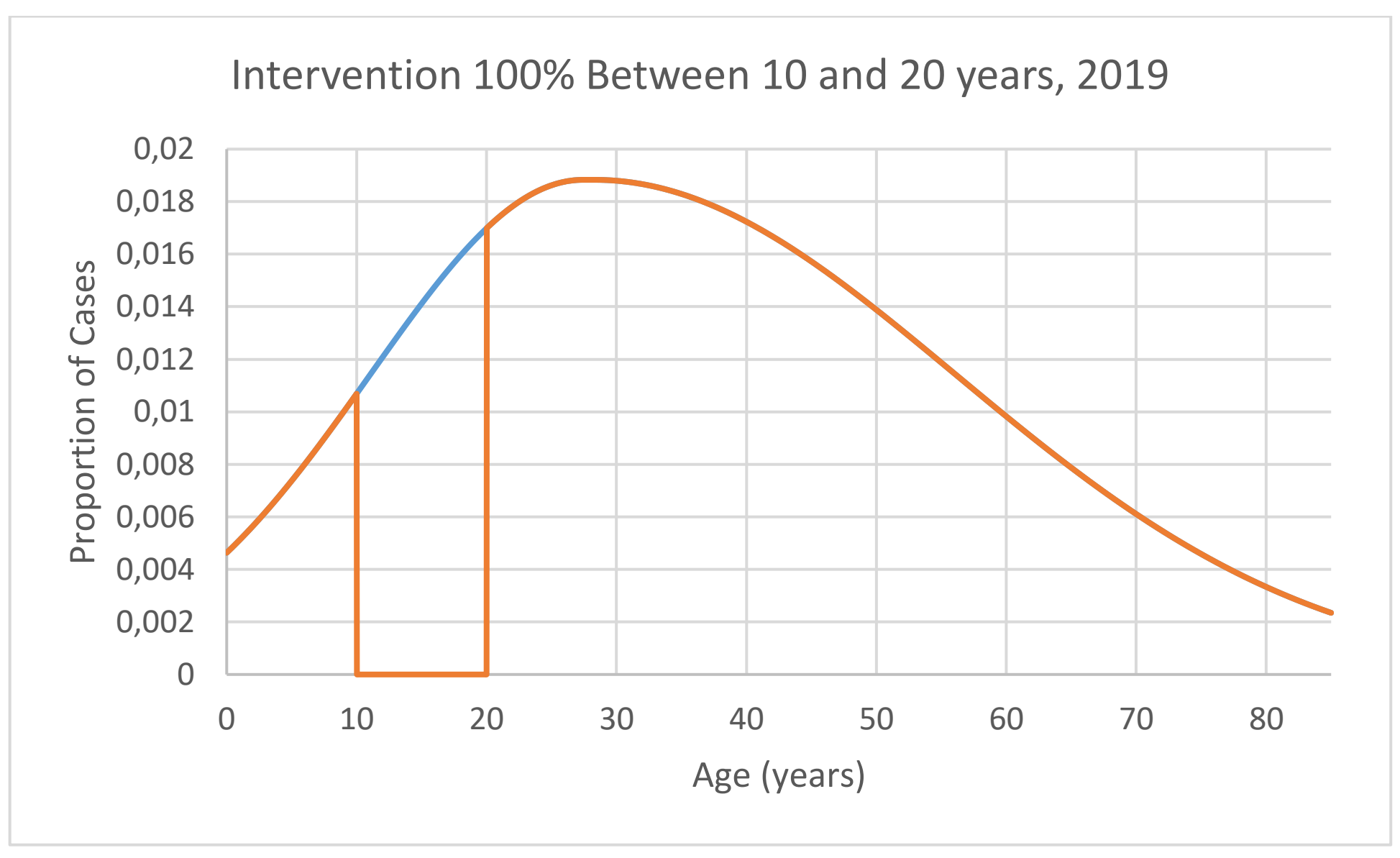

**Figure 3. Age distribution of the proportion of dengue cases in São Jose do Rio Preto in 2019. Blue line represents the observed distribution and orange line the simulated impact of an intervention between 10 and 20 years of age with $p = 100\%$.**

The number of infected vectors, in turn, in a given period of time (year) is proportional to the number of host cases in the same period of time (year). Therefore, the relationship between the number of infective vectors during the outbreak and before the intervention, $\mathcal{I}_{MT}^{before}$, and after, $\mathcal{I}_{MT}^{after}$, an intervention, say a vaccine, is given by

$$\mathcal{I}_{MT}^{after} = \alpha(a_1, a_2)\mathcal{I}_{MT}^{before}, \quad \alpha(a_1, a_2) < 1 \tag{11}$$

where $\alpha(a_1, a_2)$ is proportional to the ratio between the two areas shown in Figure 3 above, as shown below. Note that, since the proportion of cases, $\phi(a)$, is independent of the year and location (Coutinho et al. 2026), $\alpha(a_1, a_2)$ is also independent of the year and location (see Coutinho et al., 2026). With the intervention illustrated in Figure 3, we obtained $\alpha(a_1, a_2) = 0.8509$.

Let us again consider an outbreak of a vector-borne infection beginning at the start of the transmission season (raining season), $t = 0$, and ending at $t = T$ (end of the raining season) and the total number of infected mosquitoes in the season ending at time $T$, $\mathcal{I}_{MT}$ defined by equation (7).

Denoting the density of susceptible mosquitoes between $t$ and $t + dt$, $\mathcal{s}_M(t)$, we have that the total number of susceptible mosquitoes involved in the outbreak is given by

$$\mathcal{S}_{MT} = \int_0^T \mathcal{s}_M(t)dt \tag{12}$$

Now, the number of infected mosquitoes, which acquired the infection between $t$ and $t + dt$ is given by:

$$\imath_M(t)dt = k_1 d\{\Lambda(t)\mathcal{s}_M(t)\}dt \tag{13}$$

Recall that $\Lambda(t)dt$ is the number of new human cases between $t$ and $t + dt$ (see equation (1)).

After integrating (13), and assuming that the outbreak begins with virtually zero cases and zero infected mosquitoes, we obtain

$$\mathcal{J}_{MT} = k_1 \mathcal{S}_{MT} \Lambda_T \tag{14}$$

where, as before $\mathcal{J}_M(T)$, is the total number of infected mosquitoes involved in the outbreak before the intervention. After the intervention, a fraction $p$ of the number of cases is 'removed' from the age interval between $a_1$ and $a_2$.

Then, equation (14), that is, the total number of infected mosquitoes, during the outbreak, before the intervention now reads

$$\mathcal{J}_{MT}^{before} = k_1 \mathcal{S}_{MT}^{before} \Lambda_T^{before} \tag{15}$$

As the intervention occurs in a fraction $p$ of the proportion of human cases in the age interval between $a_1$ and $a_2$, we have:

$$\Lambda_T^{after} = \Lambda_T^{before} \left( 1 - \frac{p \int_0^T \left[ \int_{a_1}^{a_2} Inc_H(a,t) da \right] dt}{\Lambda_T^{before}} \right) \tag{16}$$

So,

$$\mathcal{J}_{MT}^{after} = k_1 \mathcal{S}_{MT}^{after} \Lambda_T^{before} \left( 1 - \frac{p \int_0^T \left[ \int_{a_1}^{a_2} Inc_H(a,t) da \right] dt}{\Lambda_T^{before}} \right) \tag{17}$$

Therefore, we can write

$$\frac{\mathcal{J}_{MT}^{after}}{\mathcal{J}_{MT}^{before}} = \frac{\mathcal{S}_{MT}^{after}}{\mathcal{S}_{MT}^{before}} \left( 1 - \frac{p \int_0^T \left[ \int_{a_1}^{a_2} Inc_H(a,t) da \right] dt}{\Lambda_T^{before}} \right) = \alpha(a_1, a_2, T) \tag{18}$$

Since the impact of the intervention on the total number of susceptible mosquitoes is negligible, that is $\frac{\mathcal{S}_{MT}^{after}}{\mathcal{S}_{MT}^{before}} \approx 1$ (as shown empirically since the Macdonald's works (1952), and by model simulation, by Coutinho et al, (2025) equation (18) reads:

$$\frac{\mathcal{J}_{MT}^{after}}{\mathcal{J}_{MT}^{before}} = \left( 1 - \frac{p \int_0^T \left[ \int_{a_1}^{a_2} Inc_H(a,t) da \right] dt}{\Lambda_T^{before}} \right) = \alpha(a_1, a_2, T) \tag{19}$$

Now, the reduction in the number of hosts from the transmission chain is given by removing the same proportion $p$ in the number of susceptible individuals in the population, in the same age interval. However, to know the exact number of susceptible individuals it is necessary to perform a seroprevalence survey in a representative sample of the population, previously to the intervention.

In order to estimate the number of susceptible individuals to dengue in SJRP, we used the results of Chiaravalloti-Neto et al. (2019), who performed a detailed age-dependent seroprevalence survey. Figure 4 shows the curves fitted to the densities of susceptible individuals, as compared to the infected and total population in SJRP in 2019.

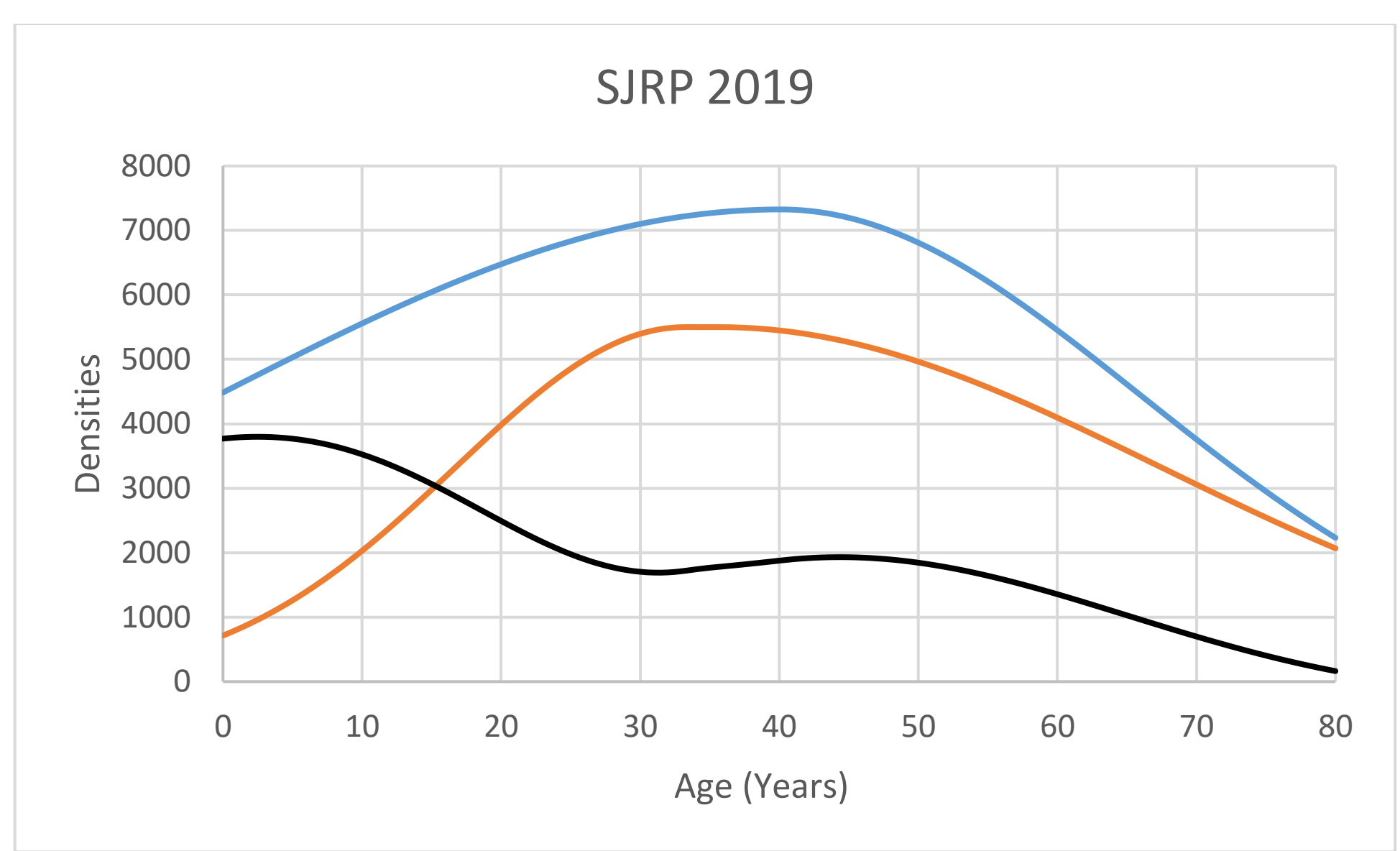

**Figure 4. Age distribution of the densities of susceptible, infected and total individuals in São José do Rio Preto in 2019. Blue line represents the distribution of the total population, orange line the infected individuals and black line the susceptible individuals in the city in 2019.**

Calling the proportion of susceptible hosts before the intervention, $\mathcal{S}_{HT}^{before}/\mathcal{N}_{HT}$ and after the intervention $\frac{\mathcal{S}_{HT}^{after}}{\mathcal{N}_{HT}}$, we have:

$$\frac{\mathcal{S}_{HT}^{after}}{\mathcal{N}_{HT}} = \beta(a_1, a_2, T)\frac{\mathcal{S}_{HT}^{before}}{\mathcal{N}_{HT}} \qquad (20)$$

where $\beta(a_1, a_2, T)$ is the ratio between the two areas shown in Figure 4 below.

Let $s_H(a,t)$ denote the total density of the age-dependent proportion of susceptible humans at time between $t$ and $t + dt$ and age between $a$ and $a + da$. Then

$$\mathcal{S}_{HT}^{before} = \int_0^T \int_0^\infty s_H(a,t) da\, dt \tag{21}$$

$$\mathcal{S}_{HT}^{after} = \int_0^T \left[ \int_0^{a_1} s_H(a,t)da + \int_{a_1}^{a_2} s_H(a,t)da + \int_{a_2}^{\infty} s_H(a,t)da - p \int_{a_1}^{a_2} s_H(a,t)da \right] dt \tag{22}$$

So

$$\frac{\mathcal{S}_{HT}^{after}}{\mathcal{S}_{HT}^{before}} = 1 - p \int_0^T \int_{a_1}^{a_2} s_H(a,T) da\, dt = \beta(a_1, a_2, T) \tag{23}$$

where the intervention occurs in a proportion $p$ of the susceptible individuals in the age interval between $a_1$ and $a_2$. However, as mentioned above, the intervention is normally carried out in the total population, due to the practical limitations to identify the exact number, or proportion, of susceptible individuals.

In Figure 5 we show the effect of an intervention of 100% in the density of susceptible individuals in the age interval between 10 and 20-year-olds.

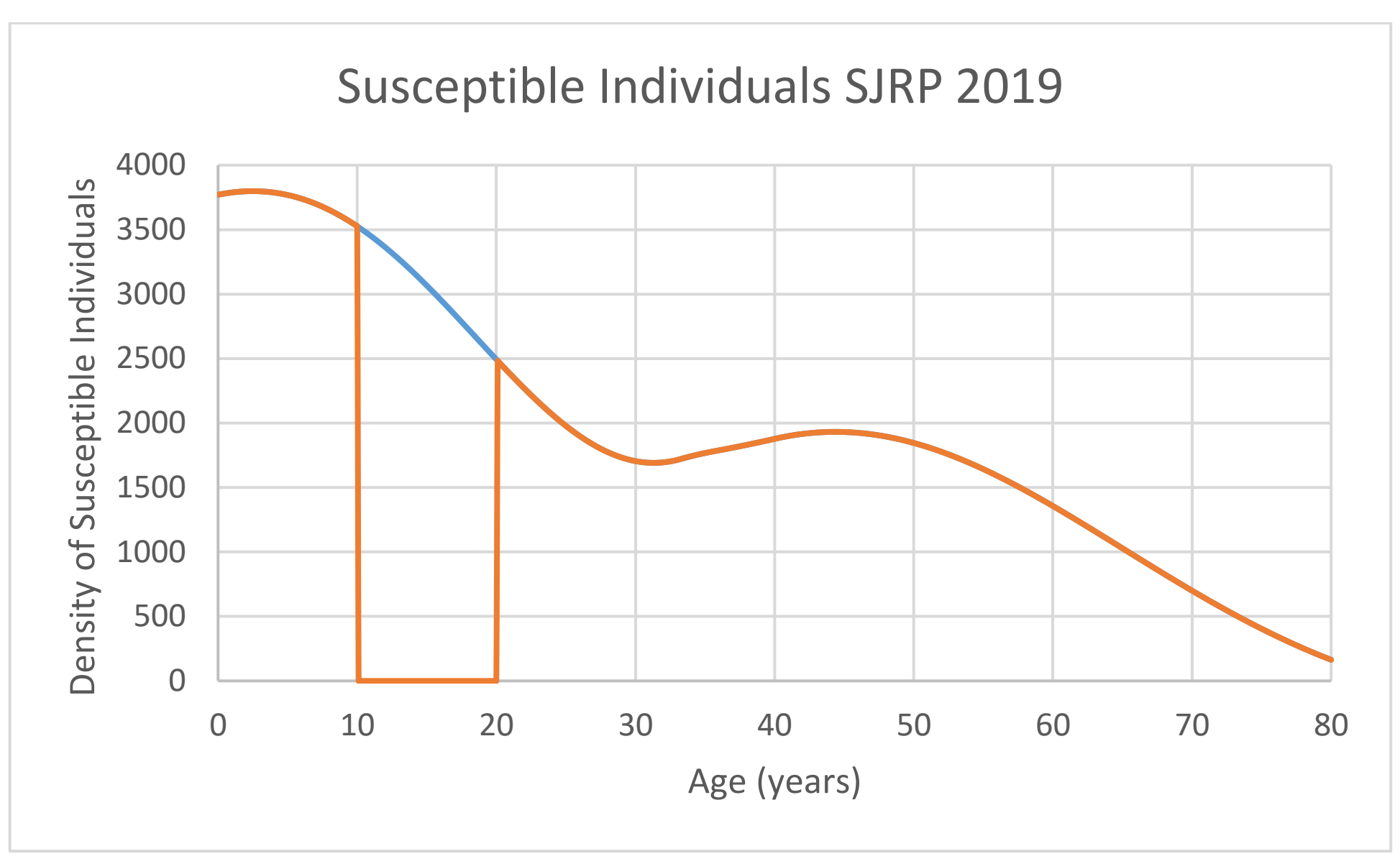

**Figure 5. Age distribution of the density of susceptible individuals in São José do Rio Preto in 2019. Blue line represents the observed distribution and orange line the simulated impact of an intervention between 10 and 20 years of age with $p = 100\%$.**

Note that, in the above simulation we assumed, as mentioned above, that the intervention was applied to the number of susceptible individuals, estimated by a seroprevalence study carried in that city (Chiravalotti-Neto et al. 2019). Although the study by Chiaravalotti-Neto et al. (2019) was carried out in 2016, the outbreaks between 2016 and 2019 were negligible, as can be seen in Figure 2.

In real situations, intervention is carried out in the entire number of individuals because the estimation of the truly susceptible is not feasible in the field. However, if we write

$$\mathcal{N}_{HT}^{before} = \mathcal{S}_{HT}^{before} + \mathcal{I}_{HT}^{before} + \mathcal{R}_{HT}^{before} \tag{24}$$

and assuming that the intervention is given by:

$$\mathcal{S}_{HT}^{after} = \mathcal{S}_{HT}^{before} - \big(1 - \beta(a_1, a_2, T)\big)\mathcal{S}_{HT}^{before} \tag{25}$$

$$\mathcal{I}_{HT}^{after} = \mathcal{I}_{HT}^{before} \tag{26}$$

$$\mathcal{R}_{HT}^{after} = \mathcal{R}_{HT}^{before} + \big(\beta(a_1, a_2, T)\big)\mathcal{S}_{HT}^{before} \tag{27}$$

Then

$\mathcal{S}_{HT}^{after} = \beta(a_1, a_2, T)\mathcal{S}_{HT}^{before}$, which is equation (18). With the intervention illustrated in Figure 4, we obtained $\beta(a_1, a_2) = 0.84724$.

Note that the intervention does not apply to the already infectious and transform the part of the susceptible into the recovered/removed.

Now the total number of cases before, $\Lambda_T^{before}$, and after, $\Lambda_T^{after}(a_1, a_2, T)$, are

$$\Lambda_T^{before} = k\mathcal{S}_{HT}^{before}\mathcal{I}_{MT}^{before} \tag{28}$$

and

$$\Lambda_T^{after}(a_1, a_2, T) = k\mathcal{S}_{HT}^{after}\mathcal{I}_{MT}^{after} \tag{29}$$

Substituting (17) and (21) in equation (27) we obtain

$$\Lambda_T^{after}(a_1, a_2, T) = k\beta(a_1, a_2, T)\mathcal{S}_{HT}^{before}\alpha(a_1, a_2)\mathcal{I}_{MT}^{before} \tag{30}$$

so that

$$\frac{\Lambda_T^{after}(a_1, a_2, T)}{\Lambda_T^{before}} = \alpha(a_1, a_2)\beta(a_1, a_2, T) \tag{31}$$

Figure 6 shows the result of the simulated intervention on the number of cases $\Lambda_T^{after}(a_1, a_2, T)$.

Therefore, the efficacy of the intervention, $Eff_T$, is (Coutinho et al., 2024)

$$Eff_T = 1 - \alpha(a_1, a_2)\beta(a_1, a_2, T) \tag{32}$$

With the intervention illustrated in Figures 2 and 3, we obtained $Eff_T = 28\%$.

Now, let us consider the age-distribution of cases before and after the intervention.

Before the intervention, we have, after re-writing equation (3):

$$\Lambda_T^{Before}(a) = \phi(a)\int_0^{\infty}\int_0^{T} Inc_H(a,t)dadt = \Lambda_T^{Before}\phi(a) \tag{33}$$

and the age-distribution of cases after is given by:

$$\Lambda_T^{After}(a) = \alpha\beta\Lambda_T^{Before}\phi_c(a, a_1, a_2, p) \tag{34}$$

where $\phi(a_1, a_2, a)$ is the orange curve in Figure 3.

Now, the result of the simulation with the same conditions described above, we obtain $\Lambda^{Before}(a)$ and $\Lambda^{After}(a)$ as shown in Figure 6.

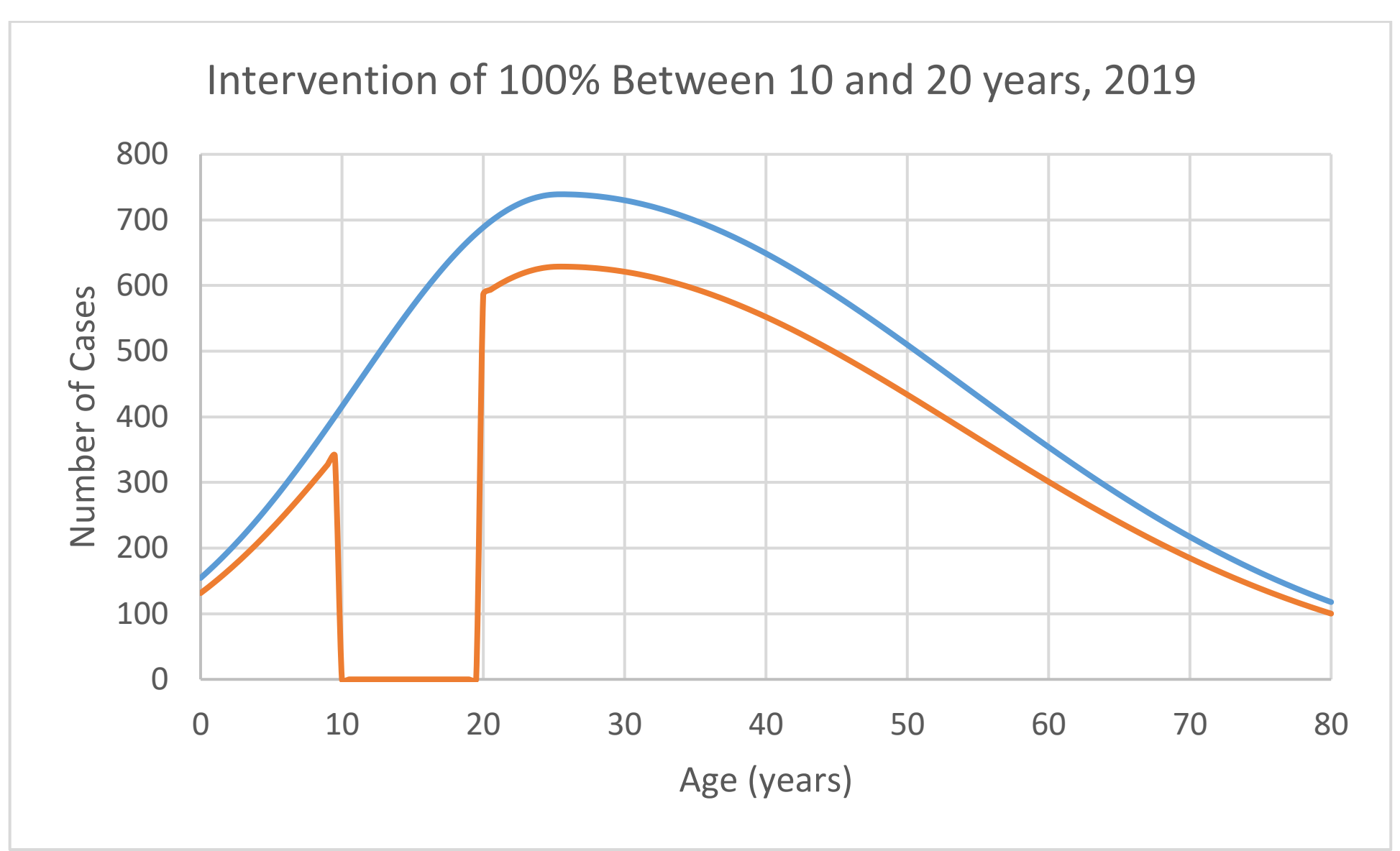

**Figure 6. Age distribution of the number of cases in São José do Rio Preto in 2025. Blue line represents the observed distribution and orange line the simulated impact of an intervention between 10 and 20 years of age with $p = 100\%$.**

In the above simulation, the efficacy of the intervention, calculated as $1 - \frac{\Lambda_T^{after}(a_1,a_2,T)}{\Lambda_T^{before}}$, resulted in 28%, which is the same value as the calculation according to equation (32).

Challenges in estimating the actual number of susceptible individuals in a real intervention when one cannot distinguish immunes from susceptible individuals.

In the previous section, we simulated an intervention in a population in which the actual number of susceptible individuals was known thanks to a previously carried out seroprevalence study (Chiaravalloti-Neto et al. 2019). The simulation was done over 100% of the susceptible fraction of the population. However, in real situations, the intervention is made blindly, covering all non-infected individuals in the age window targeted. Therefore, to guarantee that we fully protect 100% of the susceptible individuals, we have to vaccinate 100% of the population, even though a fraction of the vaccines is wasted in immune individuals.

Suppose, however, that we succeed to vaccinate only 50% of the population in the target age window. How many susceptible individuals did we really protect? The answer to this question is crucial to the calculation of parameter $\beta(a_1, a_2, T)$ and the intervention efficacy through equation (32).

If we denote $\mathcal{S}_s$ the number of susceptible individuals in a sample $\mathcal{n}$ of the population $\mathcal{N}$, the proportion of susceptible individuals in the sample are distributed according to a hypergeometric distribution $\mathcal{H}(\mathcal{N},\mathcal{S}_s,\mathcal{n})$ (Feller, 1968). For instance, for the age interval 10-19 year-olds in SJRP, Chiaravalloti-Neto et al. (2019), found $\mathcal{S}_s = 10{,}235$ in $\mathcal{N} = 27{,}738$ individuals. The probability distribution of susceptibility in that age interval resulted in an expected value for sample proportion, $\hat{p}$, of $\mathbb{E}(\hat{p}) = 5{,}118$ and standard deviation $SD(\hat{p}) = 40.2$. In this case, if we vaccinate 5,118 individuals we protect between 5,039 and 5,197 individuals. In contrast, to protect 100% of the susceptible individuals, we should vaccinate the entire population. Using these last values we could calculate the errors associated with the efficacy.

## Simulating an intervention with a real vaccine (Butantan-DV)

In this section we simulate what would be the impact of a recently licensed vaccine – Butantan-DV developed by the Instituto Butantan (Peixoto de Miranda et al. 2025) in São José do Rio Preto for the 2019 dengue outbreak.

According to the fabrication leaflet, this vaccine should be applied for the age-interval 12-59-year-olds and have a 74.7% general efficacy.

The intervention with this vaccine was simulated for age intervals with different effective immune protection. For instance, to obtain 60% of effective immune protection with a 75% effective vaccine, one should vaccinate 79% susceptible individuals between 12 and 59-year-olds (see cell in boldface in Table 1). However, as mentioned above, in practice the vaccination is applied to all the individuals complying with the including criteria. Table 1 and Figure 7 show what would be the result of several hypothetical interventions.

**Table 1. Immune protection with a hypothetical vaccine with 75% efficacy vaccine**

| Protection/Age intervals | 12-20 | 12-25 | 12-30 | 12-35 | 12-40 | 12-45 | 12-50 | 12-55 | 12-59 |
|---|---|---|---|---|---|---|---|---|---|
| 10% | 0.03 | 0.05 | 0.07 | 0.1 | 0.12 | 0.14 | 0.16 | 0.18 | 0.19 |
| 20% | 0.06 | 0.1 | 0.14 | 0.18 | 0.23 | 0.27 | 0.3 | 0.34 | 0.36 |
| 30% | 0.09 | 0.15 | 0.21 | 0.27 | 0.32 | 0.38 | 0.43 | 0.47 | 0.5 |
| 40% | 0.12 | 0.19 | 0.27 | 0.34 | 0.41 | 0.48 | 0.53 | 0.58 | 0.62 |
| 50% | 0.14 | 0.24 | 0.33 | 0.41 | 0.49 | 0.56 | 0.63 | 0.68 | 0.71 |
| 60% | 0.17 | 0.28 | 0.37 | 0.47 | 0.56 | 0.64 | 0.7 | 0.75 | 0.79 |

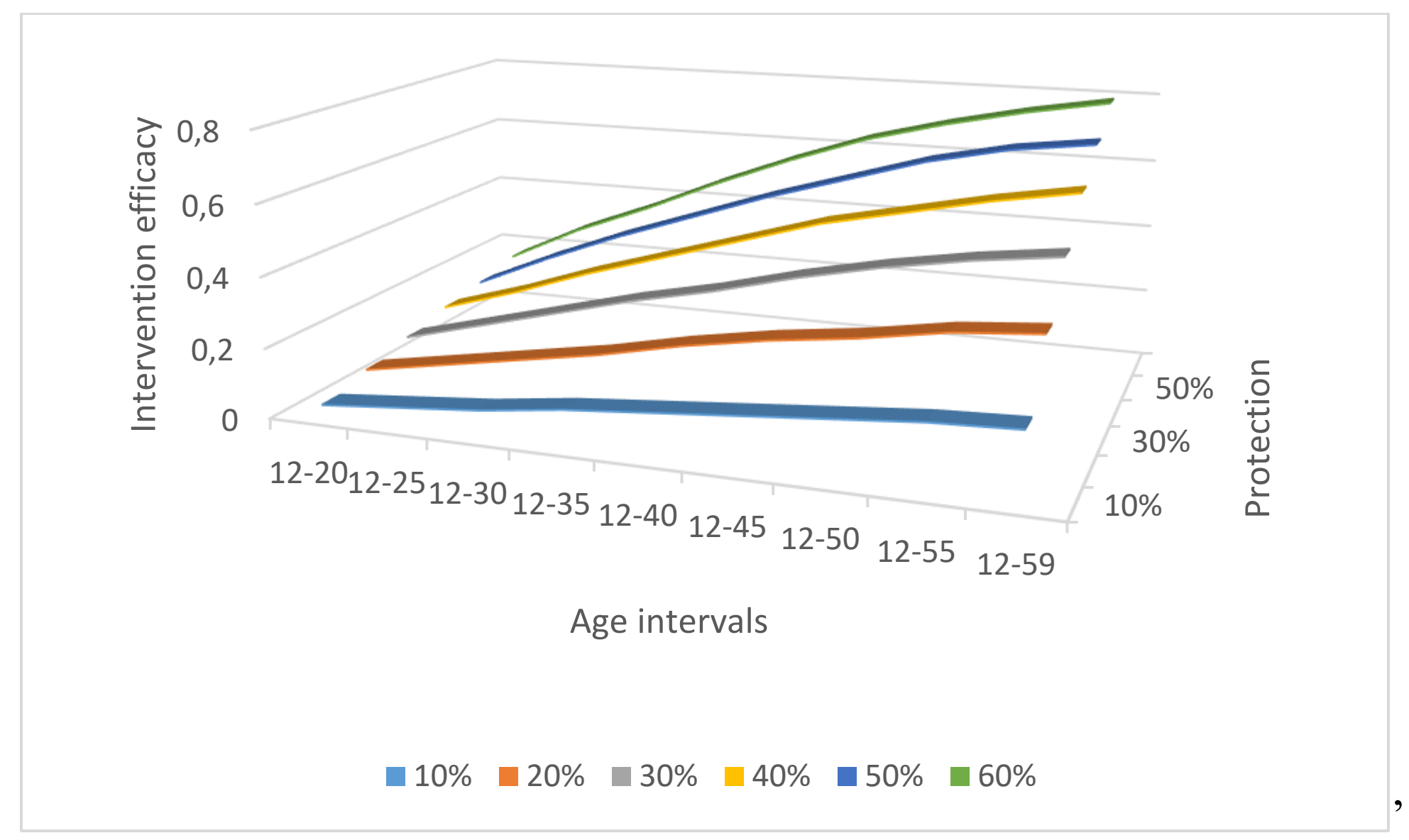

,

**Figure 7. Result of the simulations of the interventions detailed in Table 1.**

## Simulating the herd-immunity threshold

In this section, we calculate what would be the herd-immunity threshold (Anderson and May, 1991) for dengue in São José do Rio Preto in 2019. For this, we first estimated what was the basic reproduction number (Anderson and May, 1991), $R_0$ for dengue in that year, using the method described in Massad et al. (2001), Lopez et al. (2002) and Liang et al. (2018). The method consists in fitting an exponential function to the beginning of an outbreak. This allows to calculate a condition for which no outbreak could occur. That is, we calculate what would be the proportion of the population that would be protected by the intervention so that no outbreak can occur, the herd-immunity threshold.

In order to do this, we write down a deterministic system of equations involving the prevalence of infection at time $t$, $I_H(t)$ where the incidence is given by equations (1) and (8) written as:

$$\frac{dI_H(t)}{dt} = \int_0^\infty Inc_H(x,t)dx - \gamma_H I_H(t) \ = abI_M(t)\frac{S_H(t)}{N_H} - \gamma_H I_H(t) \qquad (35)$$

where, as mentioned after equation (8), $a$ is the vectors' yearly biting rate and $b$ is the fraction of the bites that result in new infections to the host, and $I_M(t)$ and $S_H(t)$ are point prevalence.

The equivalent equation for infected mosquitoes is:

$$\frac{dI_M(t)}{dt} = \exp(-\mu_M\tau)\, ac\frac{I_H(t)}{N_H}S_M(t-\tau) - \mu_M I_M(t) \qquad (36)$$

where $c$ is the fraction of the bites that result in new infections to the vectors and $\tau$ is the extrinsic incubation period of the virus inside the vector.

At the beginning of the outbreak, we can assume that the proportion of susceptible individuals (hosts and vectors), $s_H \sim 1$ and $s_M(t-\tau) \sim 1$, to get the linearized system:

$$\frac{di_H(t)}{dt} = abi_M(t) \ - \gamma_H i_H(t) \qquad (37)$$

$$\frac{di_M(t)}{dt} = \exp(-\mu_M\tau)\, aci_H - \mu_M i_M(t) \qquad (38)$$

The solution of the linearized system (35) and (36) is:

$$i_H = i_H(0)\exp(\lambda t)$$

$$i_M = i_M(0)\exp(\lambda t) \qquad (39)$$

Taking the derivative of equations (39) and substituting in equations (37) and (38), considering only the positive term, we get:

$$R_0 = \left[1 + \frac{\lambda^2 + \lambda(\mu_M + \gamma_H)}{\mu_M\gamma_H}\right]e^{\lambda\tau} \qquad (40)$$

Parameter $\lambda$ can be obtained from real data by fitting a continuous function to the initial growing phase of the outbreak, as shown in Figure 8.

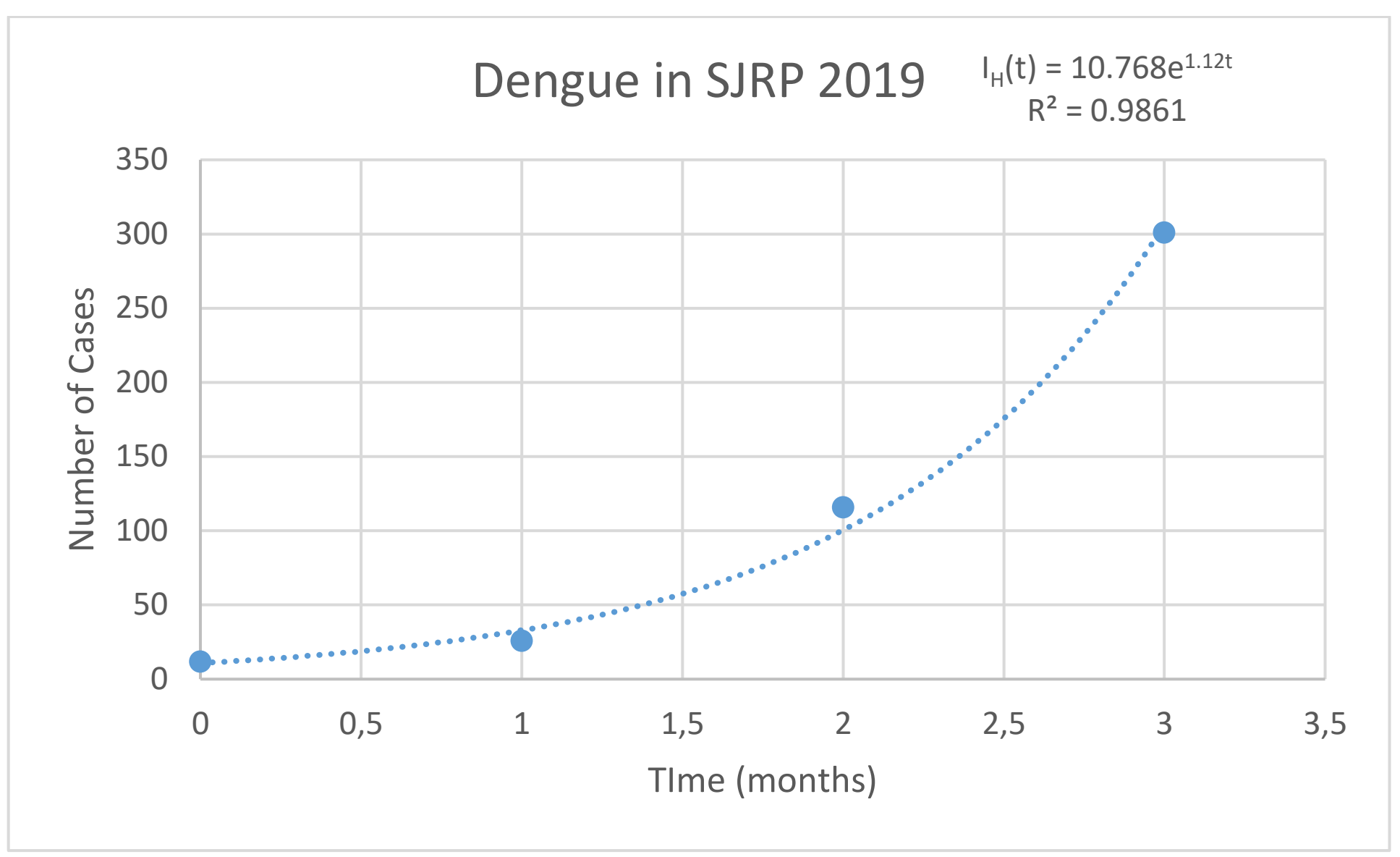

**Figure 8. Initial growing phase of the dengue outbreak in SJRP in the transmission season of 2019**

In this case $\lambda = 1.12$ and $R_0 = 5.76$.

The herd immunity threshold or critical proportion to immunize $p_c$ is related to $R_0$ by the expression (Anderson and May, 1991):

$$p_c = \left(1 - \frac{1}{R_0}\right) \tag{41}$$

Note that this critical proportion to be immunized should be corrected by multiplying it by the inverse of the vaccine efficacy, $VE$, that is $p_c \times \frac{1}{VE}$. In the example above, it results that 83% of the susceptible individuals should be vaccinated to eliminate the infection.

Numerically simulating our model to obtain the above threshold, we verify, as shown in Figure 9 that the infections were indeed eliminated.

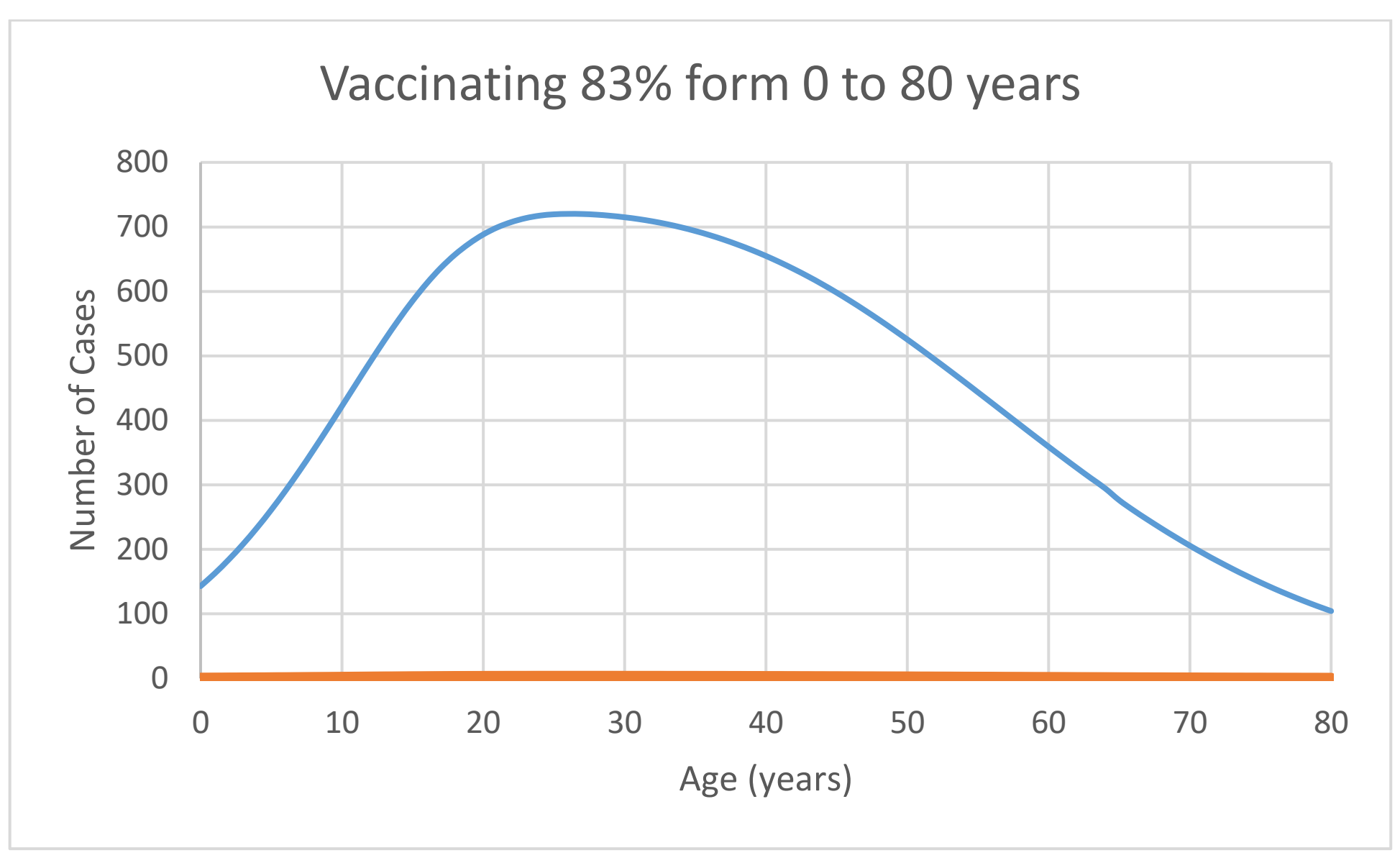

**Figure 9. Simulating the intervention at the herd immunity threshold, SJRP 2019.**

## Discussion

In this paper, we present a new method to evaluate or to design, an intervention (e.g. a vaccine) on the human population for the case of an infection to reduce the magnitude of an outbreak. The method compares what would be the number of cases due to an outbreak with and without the intervention in the same period.

For the case of a vector-borne infection, the model takes as input the officially reported age-dependent number of cases. It is deterministic and therefore does not account for any stochasticity. Our objective is to estimate the impact of the intervention (the efficacy), and we rely on the observed fact that the age distribution of the proportion of cases of the infections is independent neither of the magnitude of the outbreak nor the location where it takes place. In fact, at least for aedes-transmitted infections, this proportion is independent of both the intensity of transmission and the geographic area studied, at least for Brazilian regions (Coutinho et al, 2026).

A hypothetical intervention is simulated using a dengue vaccine, which allows the determination of the optimal strategy for a vaccination campaign. The intervention was simulated for the city of São José do Rio Preto in the State of São Paulo, a medium size city (around 490,000 inhabitants), which has a long history of high incidences of dengue and also has a reliable notification system of cases.

As mentioned in the main text, it is very difficult to predict the size of a yearly transmission season outbreak, that is, if it will be a big or a small outbreak. However, our method works for any outbreak size since it compares the results of the size of the outbreak under the intervention with the size of the outbreak without intervention.

Actually, the magnitude of the outbreak and the location in which the intervention takes place is are irrelevant. This is valid with the reservation that the above analysis is valid for medium to small size cities, because the outbreaks normally occur in different neighborhoods of large cities and the infection 'travels' along the city carried by infected humans (Amaku et al., 2016).

Some important limitations of our approach are worth mentioning. The first is that, the method demands that the hypotheses on which it rests depends on reliable notification data systems. This means that quantifying the intensity of an outbreak is dependent on the quality of the official reporting system. Therefore, biases in the notification data may compromise the quality of our conclusions.

The second important limitation is that to exemplify our method, we used a disease that is caused by four different viral strains, namely dengue. Ideally, one should choose a single-caused infection like chikungunya. However, the historical series of chikungunya is not as long as dengue. In addition, the magnitude of chikungunya outbreaks tends to be less intense, which may introduce uncertainties in the number of susceptible individuals, a variable essential to our analyses. As mentioned above, in a previous paper (Coutinho et al., 2026), however, we have shown that the age distribution of notified cases is the same for all diseases transmitted by the same vector.

The third and perhaps the most important limitation is that the method demands the knowledge of the number or proportion of susceptible humans to the disease under intervention. This is actually the Achilles heel of any method dealing with the estimation of the impact of (any) intervention. In fact, as mentioned before, the only reliable way of knowing the number or proportion of susceptible individuals in a given locality is through seroprevalence surveys, which are rarely performed due to the price and logistical hurdles involved. This is necessary, because according to our method, to estimate the effectiveness of an intervention, we have to calculate the parameter $\beta(a_1, a_2, T)$ (equation (30)). If, as in the real world, one vaccinates the whole target population it is not possible to know who has been protected. In fact, this is the limitation of any method aiming to estimate the impact of any intervention. Unless a seroprevalence test is performed before and after vaccination, the actual number of protected individuals is not known.

The intervention efficacy calculated in our example through equation (30) or by the risk ratio $(1 - \frac{\Lambda_T^{after}(a_1,a_2,T)}{\Lambda_T^{before}})$ are, as expected, the same, in spite of some approximations involved in our calculations.

To be applied to other infections, the method requires that the distribution of cases with respect to age to be determined. This is necessary to evaluate what the best age interval should be chosen to maximize the efficacy of the intervention.

Finally, it is practically impossible to verify empirically the impact of any intervention on the intensity of dengue transmission, except in the context of randomized, double blind, placebo-controlled clinical trials (Coutinho et al, 2024). Our method suffers from this limitation but this is common to any conceivable method of control of infections that do not show regular amplitude in the periodic outbreaks like dengue, as shown in Figure 2 and, in more detail, in Coutinho et al. (2026).

In conclusion, our method allows the estimation of the impact of an intervention (the efficacy) against a vector-borne infection when a seroprevalence survey is performed to determine the true number, or proportion of susceptible individuals.

**Acknowledgements** This study was supported by Fundação Butantan and CNPq.

**Funding** This work had no funding.

## Declarations

**Conflict of interest** The authors declare that they have no competing interests and that the sponsors had no role in the study design, data collection and analysis, decision to publish, or preparation of the manuscript.

**References**

Amaku M, Coutinho FAB, Azevedo RS, Burattini MN, Lopez LF, Massad E (2003) Vaccination against rubella: analysis of the temporal evolution of the age-dependent force of infection and the effects of different contact patterns. Phys Rev E. Stat Nonlin Soft Matter Phys. 67:051907. https://doi.org/10.1103/PhysRevE.67.051907

Amaku M, Coudeville L, Massad E (2012) Designing a vaccination strategy against dengue. Rev. Inst. Med. Trop. Sao Paulo 54 (Suppl 18): S18–21.

Amaku M, Azevedo F, Burattini MN, Coutinho FA, Lopez LF, Massad E (2015) Interpretations and pitfalls in modelling vector-transmitted infections. Epidemiol Infect Jul;143(9):1803-15. https://doi.org/10.1017/S0950268814002660

Amaku M, Azevedo F, Burattini MN, Coelho GE, Coutinho FAB, Greenhalgh D, Lopez LF, Motitsuki RS, Wilder-Smith A, Massad E (2016) Magnitude and frequency variations of vector-borne infection outbreaks using the Ross-Macdonald model: explaining and predicting outbreaks of dengue fever. Epidemiol Infect 144(16):3435-3450. https://doi.org/10.1017/S0950268816001448

Anderson RM, May RM (1983) Vaccination against rubella and measles: quantitative investigations of different policies. J Hyg Cam 90:259-325.

Anderson RM, May RM (1991) Infectious Diseases in Humans. Oxford University Press, Oxford.

Bauch CP, Galvani AP, Earn DJD (2003) Group interest versus self-interest in smallpox vaccination policy. Proc Natl Acad Sci USA 100 (18):10564-10567. https://doi.org/10.1073/pnas.1731324100

Boccia TM, Burattini MN, Coutinho FA, Massad E (2014) Will people change their vector-control practices in the presence of an imperfect dengue vaccine? Epidemiol Infect. 142(3):625-33. https://doi.org/10.1017/S0950268813001350

Cecílio ABI, Campanelli ES, Souza KPR, Figueiredo IB, Resende MC (2009) Natural vertical transmission by Stegomyia albopicta as dengue vector in Brazil. Braz J Biol 69 (1). https://doi.org/10.1590/S1519-69842009000100015

Chiaravalloti-Neto F, Silva RA, Zini N, Silva GC, Silva NS, Parra MCP, Dibo MR, Estofolete CF, Fávaro EA, Dutra KR, Mota MTO, Guimarães GF, Terzian ACB,

Blangiardo M, Nogueira ML (2019) Seroprevalence for dengue virus in a hyperendemic area and associated socioeconomic and demographic factors using a cross-sectional design and a geostatistical approach, state of São Paulo, Brazil. BMC Infectious Diseases 19:441. https://doi.org/10.1186/s12879-019-4074-4

Coutinho FA, Burattini MN, Lopez LF, Massad E (2006) Threshold conditions for a non-autonomous epidemic system describing the population dynamics of dengue. Bull Math Biol 68(8):2263-82. https://doi.org/10.1007/s11538-006-9108-6

Coutinho FAB, Amaku M, Boulos FC, de Sousa Moreira JA, Dias Franca JI, do Amaral JA, de Barros ENC, Struchiner CJ, Kallas EJ, Massad E. Analysing vaccine efficacy evaluated in phase 3 clinical trials carried out during outbreaks. Infect Dis Model. 2024 May 23;9(4):1027-1044. doi: 10.1016/j.idm.2024.05.007.

Coutinho FAB, Amaku M, Boulos FC, de Sousa Moreira JA, de Barros ENC, Kallas EG, Massad E. Estimating the Size of the Aedes Mosquitoes' Population Involved in Outbreaks of Dengue and Chikungunya Using a Mathematical Model. Bull Math Biol. 2025 Aug 8;87(9):124. http://doi.org/10.1007/s11538-025-01489-z

Coutinho FAB, Amaku M, Kallas EG and Massad E (2026). Age-dependent distribution of officially reported cases of vector-borne infections arXiv:2603.16773 [q-bio.PE] (or arXiv:2603.16773v1 [q-bio.PE] for this version) https://doi.org/10.48550/arXiv.2603.16773

Feller W (1968) An Introduction to Probability Theory and Its Applications (3rd ed., Vol. 1). Wiley, New York.

Ferreira CP, Lyra SP, Azevedo F, Greenhalgh D, Massad E (2017) Modelling the impact of the long-term use of insecticide-treated bed nets on Anopheles mosquito biting time, Malar. J. 16 (1) (2017) 373. http://doi.org/10.1186/s12936-017-2014-6

Guzman M, Kouri G (2002) Dengue: an update. Lancet Infect Dis 2(11):33–42. http://doi.org/10.1016/s1473-3099(01)00171-2

Hethcote HW (1988) Optimal ages of vaccination for measles. Math. Biosc 89: 29-52.

Liang Y, Ahmad Mohiddin MN, Bahauddin R, Hidayatul FO, Nazni WA, Lee HL, Greenhalgh D (2019) Modelling the effect of a novel autodissemination trap on the spread of dengue in Shah Alam and Malaysia. Comput Math Methods Med 2019:1923479. https://doi.org/10.1155/2019/1923479

Lopez LF, Coutinho FA, Burattini MN, Massad E (2002) Threshold conditions for infection persistence in complex host-vectors interactions. C R Biol 325(11):1073-84. http://doi.org/10.1016/s1631-0691(02)01534-2

Macdonald G (1952) The analysis of the sporozoite rate. Trop Dis Bull 49(6):569-86.

Massad E, Burattini MN, Azevedo Neto RS, Yang HM, Coutinho FAB and Zanetta DMT (1994) A model-based design of a vaccination strategy against rubella in a non-immunized community of São Paulo, Brazil. Epidemiol Infect 112:579-594.

Massad E, Coutinho FA, Burattini MN, Lopez LF (2001) The risk of yellow fever in a dengue-infested area. Trans R Soc Trop Med Hyg 95(4):370-4. https://doi.org10.1016/s0035-9203(01)90184-1

Massad E, Coutinho FA, Burattini MN, Lopez LF, Struchiner CJ (2005) Yellow fever vaccination: how much is enough? Vaccine 23(30):3908-14. https://doi.org/10.1016/j.vaccine.2005.03.002

Massad E, Tan SH, Khan K, Wilder-Smith A (2016) Estimated Zika virus importations to Europe by travellers from Brazil. Glob Health Action 9:31669. https://doi.org/10.3402/gha.v9.31669

Massad E, Amaku M, Coutinho FAB, Struchiner CJ, Lopez LF, Wilder-Smith A, Burattini MN (2017) Estimating the size of Aedes aegypti populations from dengue incidence data: Implications for the risk of yellow fever outbreaks. Infect Dis Model 2(4):441-454. https://doi.org/10.1016/j.idm.2017.12.001

Peixoto de Miranda ÉJF, de Sousa Moreira JA, da Silva Braga R, Silveira DHR, Infante V, de Oliveira Alves LB, de Camargo Vieira Tenorio J, Dos Santos Silva GF, Patiño EG, de Mesquita Pacheco PHT, Ramos F, Oliveira DS, Kallás EG, Boulos FC (2025) Randomized, double-blind, placebo-controlled, phase 3 trial to demonstrate lot-to-lot consistency of 3 lots of the simplified formulation of Butantan-dengue vaccine. Vaccine. 66:127836. https://doi.org/10.1016/j.vaccine.2025.127836

Pinto E, Coelho M, Oliver L, Massad E (2011) The influence of climate variables on dengue in Singapore. Int J Environ Health Res 21(6):415-26. https://doi.org/10.1080/09603123.2011.572279

Powell JR, Tabachnick WJ (2013) History of domestication and spread of *Aedes aegypti* - A Review. Mem Inst Oswaldo Cruz 108(Suppl 1):11-17. https://doi.org/10.1590/0074-0276130395

SINAN-MOH, Sistema Nacional de Agravos de Notificação, Ministério da Saúde do Brasil (2025) https://datasus.saude.gov.br/acesso-a-informacao/doencas-e-agravos-de-notificacao-de-2007-em-diante-sinan/